\documentclass[aps,prd,12point,nofootinbib,superscriptaddress]{revtex4-2}

\usepackage{color}
\usepackage{amssymb, bm}
\usepackage{amsmath, amsthm}
\usepackage{epstopdf}
\usepackage{hyperref}

\newcommand{\be}{\begin{equation}}
\newcommand{\ee}{\end{equation}}

\def\theequation{\arabic{section}.\arabic{equation}}
\begin{document}


\title{First-order thermodynamics of Horndeski gravity}

\author{Andrea Giusti}
\email[]{agiusti@phys.ethz.ch}
\affiliation{Institute for Theoretical Physics, ETH Zurich,
Wolfgang-Pauli-Strasse 27, 8093, Zurich, Switzerland}

\author{Stefan Zentarra}
\email[]{szentarra@phys.ethz.ch}
\affiliation{Institute for Theoretical Physics, ETH Zurich,
Wolfgang-Pauli-Strasse 27, 8093, Zurich, Switzerland}

\author{Lavinia Heisenberg}
\email[]{laviniah@phys.ethz.ch}
\affiliation{Institute for Theoretical Physics, ETH Zurich,
Wolfgang-Pauli-Strasse 27, 8093, Zurich, Switzerland}

\author{Valerio Faraoni}
\email[]{vfaraoni@ubishops.ca}
\affiliation{Department of Physics \& Astronomy, Bishop's University, 
2600 College Street, Sherbrooke, Qu\'ebec, Canada J1M~1Z7 }



\begin{abstract}

We extend to the Horndeski realm the irreversible thermodynamics 
description of gravity previously studied in ``first generation'' 
scalar-tensor theories. We identify a subclass of Horndeski theories 
as an out-of--equilibrium state, while general relativity corresponds 
to an equilibrium state. In this context, we identify an effective heat 
current, ``temperature of gravity'', and shear viscosity in the space of 
theories. The identification is accomplished by recasting the field 
equations as effective Einstein equations with an effective dissipative 
fluid, with Einstein gravity as the equilibrium state, following Eckart's 
first-order thermodynamics.

\end{abstract}
\maketitle

\section{Introduction}
\label{sec:1}
\setcounter{equation}{0}

A connection between gravity and thermodynamics has been drawn with the 
discovery of black hole and horizon thermodynamics and has been augmented 
by the suggestion that gravity may not be fundamental after all, but 
rather that it emerges from more fundamental constituents in the same way 
that a fluid emerges from its atoms or molecules.  An 
indication that this idea may not be too far fetched is the derivation of 
the Einstein field equation of general relativity (GR) by Jacobson as an 
equation of state, using thermodynamical considerations 
\cite{Jacobson:1995ab}. This view of gravity has profound implications for 
quantum gravity as well, since it implies that quantizing the Einstein 
equation is rather meaningless---the analogy would be that it amounts to 
quantizing the macroscopic ideal gas equation of state. Such a 
quantization could, at most, find phonons but certainly not results as 
fundamental as the eigenfunctions and energy spectrum of the Hamiltonian 
of the hydrogen atom. If they exist, the ``atoms of spacetime'' would not 
be directly related to the Einstein equation.

One then wonders what role extensions of GR may play 
in the broader context of a gravity-thermodynamics connection. A bold idea 
was advanced in the second seminal paper~\cite{Eling:2006aw}, in which the 
field equation of metric $f( R) $ gravity was also obtained with 
thermodynamical considerations. According to \cite{Eling:2006aw}, this 
fourth order modification of GR corresponds to a non-equilibrium state in 
a ``thermodynamics of gravitational theories''. Although the latter is not 
fully developed, a ``bulk viscosity of spacetime'' is introduced to drive 
a dissipation process of gravity towards a thermodynamic equilibrium 
state, which is Einstein's gravity as previously indicated in 
\cite{Jacobson:1995ab}. In other words, GR would be an equilibrium state 
while $f(R)$ gravity is an excited (or non-equilibrium) state. By 
extension, any gravity theory containing extra dynamical degrees of 
freedom with respect to the two massless spin-two modes of GR should 
correspond to an ``excited state''. Indeed, referring to extra dynamical 
degrees of freedom makes the idea of GR as a ``ground state'' appear very 
reasonable.

Extensions of gravity are well-motivated from the theoretical point of view. It 
is well known that virtually any attempt to introduce quantum corrections 
to GR involves extending it by introducing higher order corrections or 
extra degrees of freedom, and one expects extensions to GR to appear as 
soon as one moves away from the lowest energy regime.  For example, one-loop 
renormalization requires the introduction of terms quadratic in the 
curvature and, to date, the most successful scenario of early universe 
inflation \cite{Planck:2015sxf,Planck2}, {\em i.e.}, Starobinsky inflation 
\cite{Starobinsky:1980te}, is based on $R^2$ corrections to the 
Einstein-Hilbert action. The low-energy limit of the bosonic string, the 
most rudimental string theory known, does not reproduce GR but gives 
instead an $\omega=-1$ Brans-Dicke theory (where $\omega$ is the 
Brans-Dicke coupling constant) \cite{Callan:1985ia,Fradkin:1985ys}.

Today, the main motivation for extended gravity theories is no doubt given by 
cosmology \cite{Heisenberg:2018vsk}. The 1998 discovery of the accelerated expansion of the universe 
made with type~Ia supernovae calls for an explanation. The standard 
$\Lambda$-Cold Dark Matter ($\Lambda$CDM) cosmological model based on GR 
requires the introduction of either an incredibly fine-tuned cosmological 
constant $\Lambda$, or of a completely {\em ad hoc} dark energy (see 
\cite{AmendolaTsujikawabook} for a discussion).  An alternative consists 
of extending gravity at large cosmological scales without introducing 
dark energy \cite{Capozziello:2003tk,Carroll:2003wy}. Many approaches to 
extended gravity theories motivated by cosmology have been researched, with $f(R)$ 
gravity probably being the most popular (see 
\cite{Sotiriou:2008rp,DeFelice:2010aj,Nojiri:2010wj}
 for reviews). Gravity is tested rather poorly at certain scales 
and in certain regimes \cite{Berti:2015itd,Baker:2014zba}, which leaves 
ample room for extended gravity theories. In view of the 
above, it is only natural to contemplate the role of extensions when 
investigating the connection between gravity and thermodynamics.

The seminal papers \cite{Jacobson:1995ab,Eling:2006aw} have been very 
influential.
However, in spite of a large literature, no light has been shed on the 
equations ruling the dissipation of gravity towards the GR equilibrium 
state and the order parameter indicating how close the non-equilibrium is 
to the GR equilibrium state has not been identified. Overall, not much 
progress has been made in this direction since the works 
\cite{Jacobson:1995ab,Eling:2006aw}. We ought to mention, however, an 
important result:  Ref.~\cite{Chirco:2010sw} identified the essential role 
played by shear viscosity, while eliminating bulk viscosity from the 
thermodynamics of spacetime picture.

Perhaps the lack of progress is due to the fact that the ideas advanced in 
Refs.~\cite{Jacobson:1995ab,Eling:2006aw} are so fundamental that they 
require at least some more advanced knowledge of the basic ingredients in 
order to be developed. In previous papers 
\cite{Faraoni:2021lfc,Mentrelli}, we proposed a more 
modest approach in a very different context, but in the same spirit. 
First, within the wide spectrum of gravity theories, we 
identified the scalar-tensor class generalizing Brans-Dicke gravity 
\cite{Brans:1961sx,Bergmann:1968ve, Nordtvedt:1968qs, Wagoner:1970vr,
Nordtvedt:1970uv} as the most suitable candidate for our new approach. 
This class contains $f( R) $ gravity as a subclass 
\cite{Sotiriou:2008rp,DeFelice:2010aj,Nojiri:2010wj}. Brans-Dicke gravity 
\cite{Brans:1961sx} and its 
scalar-tensor generalizations \cite{Bergmann:1968ve, Nordtvedt:1968qs, 
Wagoner:1970vr,Nordtvedt:1970uv} are minimal extensions of GR 
since they contain only a scalar degree of freedom $\phi$ in addition to 
the usual two massless spin two modes of GR. The description of 
scalar-tensor gravity as an excited state of GR does not require 
fundamental assumptions such as those of 
Refs.~\cite{Jacobson:1995ab,Eling:2006aw} in spacetime 
thermodynamics (for instance, the notions of local causal horizon and 
of local Rindler frame to take advantage of a local Unruh effect). In fact, 
the structure of a dissipative fluid containing GR as its limit is already 
contained in the field equations of scalar-tensor gravity. More precisely, 
the contribution of the scalar degree of freedom $\phi$ to the field 
equations has the structure of the energy-momentum tensor of  an 
effective relativistic dissipative 
fluid \cite{Pimentel89,Faraoni:2018qdr}. This fact can be 
derived in a straightforward manner by rewriting the field equations and 
is not an independent assumption. 

By taking seriously the dissipative fluid nature of the effective 
stress-energy tensor of $\phi$, one wonders what the minimal theory of 
relativistic dissipative fluids has to say. With this question in mind, we 
have applied Eckart's first-order thermodynamics 
\cite{Eckart:1940te,Maartens:1996vi,Andersson:2006nr} (which is non-causal 
and 
plagued by instabilities, but is nevertheless the most widely applied 
formalism to describe dissipation in GR)  to the effective $\phi$-fluid. 
Explicit expressions for the corresponding thermodynamical quantities were 
obtained, including the heat current density, the sought-for ``temperature 
of extended gravity'', and shear viscosity.

In the last decade, ``first generation'' scalar-tensor gravity has been 
generalized by rediscovering and reformulating \cite{H1,H2,H3} the old 
Horndeski gravity \cite{Horndeski}, which still contains only an extra 
scalar degree of freedom obeying second order equations of motion, but is 
much more general. These efforts were followed by the discovery that 
higher order Lagrangians still admit second order equations of motion and 
are healthy with respect to the Ostrogradski instability when subjected to 
a degeneracy condition, resulting in the so-called Degenerate Higher Order 
Scalar-Tensor (``DHOST'') theories \cite{GLPV1,GLPV2}. These 
generalizations, developed in \cite{DHOST1, DHOST3, DHOST3, DHOST4, 
DHOST5, DHOST6, DHOST7}, introduce a large number of terms in the 
gravitational sector of the action and have generated a rich literature 
(see Refs.~\cite{DHOSTreview1, DHOSTreview2} for reviews). DHOST theories 
are restricted by theoretical constraints avoiding graviton decay into 
scalar field perturbations \cite{Creminellietal18} and, above all, by the 
recent multi-messenger observation of simultaneous gravitational waves and 
$\gamma$-rays in the GW170817/GRB170817 event 
\cite{TheLIGOScientific:2017qsa,Monitor:2017mdv}. The latter sets severe 
constraints on the space of DHOST and Horndeski theories from the 
observed upper limits on the difference between the propagation speeds 
of gravitational and electromagnetic waves \cite{Langloisetal18}.

Motivated by the explosion of interest in Horndeski gravity, we apply the 
effective fluid formalism which was successul in ``first generation'' 
scalar-tensor gravity to this more general scenario. In particular, 
we rewrite the field equations as effective Einstein equations containing 
an effective dissipative fluid in their right hand side (at this stage, we 
complete and correct the previous work \cite{Quiros:2019gai} which 
computes some of the effective fluid quantities for a subclass of 
Horndeski theories, but does not work out the thermodynamics of the 
corresponding dissipative fluid). Then, we proceed to identifying the 
corresponding heat flux density, anisotropic stresses, ``temperature of 
gravity'', and shear viscosity whenever possible.

As a result, the effective thermodynamics of scalar-tensor gravity does 
not apply to the most general Horndeski theory because some 
terms in the field equations explicitly break the thermodynamic analogy
by spoiling the constitutive equation of Eckart's theory  for the 
effective $\phi$-fluid,  
contrary to the first generation scalar-tensor theories examined in 
\cite{Faraoni:2021lfc,Mentrelli}. However, this situation returns to 
``normal'' when the most general Horndeski action is restricted by 
eliminating the terms violating the equality between the propagation 
speeds of gravitational and electromagnetic waves. In this sense, the 
thermodynamics of gravity seems to indicate the way to the physical 
constraints on Horndeski gravity. Notably, the terms leading to the failure of
the thermodynamic analogy are those operators which contain non-minimal derivative couplings
and non-linear contributions in the connection. 
These operators are exactly the ones that do not allow a local field redefinition into the Jordan frame 
and give rise to intrinsic modifications of the gravity sector. 
It is intriguing that the analogy between the effective fluid description of scalar-tensor theories and 
Eckart's first-order thermodynamics is tightly related to these intrinsic changes of the helicity-2 sector. 
However, since the failure of rewriting the theory in the Jordan frame is directly related to the failure of separating
the ``gravity effects" from the ``effective matter fluid'', this result does not come as a surprise.
 
We follow the notation of Ref.~\cite{Waldbook}: the metric signature is 
$({-}{+}{+}{+})$ and units are used in which the speed of light $c$ and 
Newton's constant $G$ are unity.


\section{Kinematic quantities for the effective fluid}
\label{sec:3}
\setcounter{equation}{0}

Assuming that the scalar field gradient is timelike, $
\nabla _a \phi \nabla ^a \phi < 0 $, and using the notation
\be
X \equiv - \frac{1}{2} \nabla_a \phi \nabla^a \phi >0  \,,
\ee
it is natural to define the 4-velocity of the effective fluid 
associated with the scalar field $\phi$ as 
\be 
u^a = \frac{\nabla ^a \phi}{\sqrt{2  X}} \, , 
\ee
which satisfies $u^a u_a = -1$. Identifying the field $u^a$ uniquely 
identifies 
a frame comoving with the effective fluid and a $3+1$ splitting of 
spacetime in 
the time direction of these observers with 4-tangent $u^a$ and their 
3-space.  
The Riemannian metric on the 3-space 
orthogonal to $u^a$ ({\em i.e.}, the 3-space of the observers comoving 
with the effective fluid) is 
\be
 h_{ab} = g_{ab} + u_a u_b = g_{ab} + \frac{\nabla _a \phi \nabla _b 
\phi}{2 X} \,,
\ee
while ${h^a}_b$ is the projection operator onto this 3-space.  The  
4-velocity gradient is
\be
\begin{split}
\nabla _a u_b & 
= \frac{1}{\sqrt{2 X}} \left( \nabla _a \nabla _b \phi - 
\frac{\nabla_a X \nabla _b \phi}{2 X} \right)  \, .
\end{split}
\ee
Using the fact  that 
\be
\nabla _a X = - \frac{1}{2} \nabla _a \left( \nabla _e \phi \nabla ^e 
\phi \right)  
= - \nabla _e \phi \nabla _a \nabla ^e \phi \, ,
\ee
this 4-velocity gradient is written as
\be
\nabla _a u_b = \frac{1}{\sqrt{-\nabla _e \phi \nabla ^e \phi}} 
\left( \nabla _a \nabla _b \phi + \frac{\nabla _b \phi \nabla _c \phi 
\nabla _a \nabla ^c \phi}{-\nabla _e \phi \nabla ^e \phi} \right) \, ,
\ee
that coincides with the corresponding expressions appearing in  
Refs.~\cite{Faraoni:2018qdr,Faraoni:2021lfc}.

The effective fluid has 4-acceleration 
\be
\label{eq:udot}
\begin{split}
\dot{u}^a & \equiv u^c \nabla_c u^a = \frac{\nabla ^c 
\phi}{\sqrt{2 X}} \frac{1}{\sqrt{2 X}}  \left( \nabla _c \nabla ^a \phi - 
\frac{\nabla_c X \nabla ^a \phi}{2 X} 
\right)\\
&= - \frac{1}{{2 X}} \left(\nabla ^a X + \frac{\nabla X \cdot 
\nabla \phi}{2 X} \nabla ^a \phi \right) \, ,
\end{split}
\ee
where $\nabla X \cdot \nabla \phi \equiv g^{ab} \nabla_a X  \nabla _b 
\phi$.

Using ${h_a}^b = {\delta _a}^b + u_a u^b$ and $u^a \nabla_b u_a = 
0$, the (double) spatial projection of the velocity gradient is 
\be
\begin{split}
V_{ab} & \equiv {h_a}^c {h_b}^d \nabla _d u_c = \nabla _b u_a + \dot{u}_a 
\, 
u_b\\
&= \frac{1}{\sqrt{2 X}} \left( \nabla _b \nabla _a \phi - \frac{\nabla_b 
X \nabla _a \phi}{2 X} \right) 
- \frac{\nabla _b \phi}{\sqrt{2 X}} \frac{1}{{2 X}} \left(\nabla _a X + 
\frac{\nabla X \cdot \nabla \phi}{2 X} \nabla _a \phi \right)\\
&= \frac{1}{\sqrt{2 X}} \left( \nabla _a \nabla_b \phi - 
\frac{\nabla_{(a} X \nabla _{b)} \phi}{X} 
- \frac{\nabla X \cdot \nabla \phi}{4 X^2} \nabla _a \phi \nabla _b \phi 
\right)  \, .
\end{split}
\ee
This tensor is symmetric, $V_{ab} = V_{(ab)}$, and its antisymmetric part 
vanishes identically,
\be
\omega_{ab} \equiv V_{[ab]} = 0 \,,
\ee
so that the 4-velocity $u^a$ of the effective $\phi$-fluid is 
irrotational\footnote{This feature was missed in Ref.~\cite{Quiros:2019gai}.} 
and hypersurface-orthogonal, the spacetime line element $ds^2=g_{ab}dx^a dx^b$ 
becomes diagonal in adapted coordinates $\big\{ x^a \big\}$, and a 
foliation of 
3-dimensional hypersurfaces with Riemannian metric $h_{ab}$ always exists 
\cite{Waldbook}. The expansion scalar of the effective fluid is
\be
\begin{split}
\theta &= \nabla _a u^a = 
\frac{1}{\sqrt{2 X}} \left( \Box \phi - \frac{\nabla X \cdot 
\nabla \phi}{2 X} \right) \\
&= \frac{1}{\sqrt{-\nabla_c \phi \nabla ^c \phi}} 
\left( \Box \phi - \frac{\nabla ^a \phi \nabla ^b \phi \nabla _a \nabla 
_b \phi}{\nabla _e \phi \nabla ^e \phi} \right) \,,
\end{split}
\ee
while the trace-free shear tensor reads 
\be
\label{eq:shear}
\begin{split}
\sigma _{ab} & \equiv V_{(ab)} - \frac{\theta}{3} \, h_{ab} = V_{ab} - 
\frac{\theta}{3} \, h_{ab}\\
&= \frac{1}{\sqrt{2 X}} \left[ \nabla _a \nabla_b \phi - 
\frac{\nabla_{(a} X \nabla _{b)} \phi}{X} 
 - \frac{\nabla X \cdot \nabla \phi}{4 X^2} \nabla _a \phi \nabla _b \phi
- \frac{h_{ab}}{3} \left( \Box \phi - \frac{\nabla X \cdot \nabla \phi}{2 
X} \right) \right] \,.
\end{split}
\ee
These kinematic quantities for the effective $\phi$-fluid coincide with 
those reported in the previous Refs.~\cite{Faraoni:2018qdr, 
Faraoni:2021lfc} for the ``old'' scalar-tensor gravities of 
\cite{Brans:1961sx,Bergmann:1968ve, Nordtvedt:1968qs, Wagoner:1970vr, 
Nordtvedt:1970uv}. This fact is expected because shear, expansion and 
vorticity are purely kinematic quantities, and cannot depend on the 
particular theory ({\em i.e.}, on the action or the field equations), 
provided that only a scalar degree of freedom $\phi$ is added to the 
ordinary spin~2 massless polarizations of the metric tensor in Einstein 
gravity, as done in Brans-Dicke or in Horndeski gravity.

\section{Imperfect fluid description of Horndeski gravity}
\label{sec:4}
\setcounter{equation}{0}

The most general Lagrangian of Horndeski gravity reads 
\be
\mathcal{L} = \mathcal{L}_2 + \mathcal{L}_3 + \mathcal{L}_4 + 
\mathcal{L}_5 \,,
\ee
where the individual interactions are given by
\begin{eqnarray}
\label{eq:HorndeskiGeneral}
 \mathcal{L}_2 &=& G_2 \, , \\
&&\nonumber\\
\nonumber \mathcal{L}_3 &=& - G_3 \, \Box \phi \, ,\\
&&\nonumber\\
\nonumber \mathcal{L}_4 &=& G_4 \, R + G_{4 X} \left[ (\Box \phi)^2 - 
(\nabla_a \nabla_b \phi)^2 \right] \,,\\
&&\nonumber\\
\nonumber \mathcal{L}_5 &=& G_5 \, G_{ab} \, \nabla^a \nabla^b 
\phi -  \frac{G_{5X}}{6} \Big[  (\Box \phi)^3   - 3 \, \Box \phi \, (\nabla_a \nabla_b \phi)^2 
+ 2 \, (\nabla_a \nabla_b \phi)^3 \Big] \,,
\end{eqnarray}
where $\phi$ is the scalar degree of freedom, $X \equiv - \nabla^c 
\phi \nabla_c \phi/2 $, $\nabla_a $ is the covariant derivative of the 
metric $g_{ab}$ (which has determinant $g$), and $\Box \equiv g^{ab} 
\nabla_a \nabla_b $ is d'Alembert's operator, $G_{ab}$ denotes the 
Einstein tensor, while $G_i (\phi, X)$ ($i=2,3,4,5$) are arbitrary functions of 
the scalar field $\phi$ and of the canonical kinetic term $X$. 
Note that, according to the standard notation, we define 
$G_{i\phi} \equiv \partial G_i / \partial  \phi$ and $G_{iX} \equiv \partial G_i / \partial  X$.

This Lagrangian represents the most general scalar-tensor theory with second order equations of motion.
The local and second order nature of the field equations then implies that the theory naturally avoids
Ostrogradsky instabilities.

Here we consider a sub-class of Horndeski gravity, namely the one that implies 
a luminal propagation of gravitational waves, in which the interactions are restricted to 
\be
\label{eq:LsH}
\mathcal{\bar{L}}= G_2 (\phi, X) - G_3 (\phi, X) \Box \phi + G_4 (\phi) R \, ,
\ee 
that is, we set $G_{4 X} = 0$, $G_5=0$. This last Lagrangian will be the main focus in the following.

Performing the variation of the above Lagrangian
\be
\delta \left( \sqrt{-g} \, \mathcal{\bar{L}}  \right) = \sqrt{-g} \, 
\sum_{i = 2}^4 
\mathcal{G} ^{(i)} _{ab} \, \delta g^{ab} + \sum_{i = 2} ^4 (\cdots) \, 
\delta \phi + \mbox{total derivative} 
\ee
and specializing the results in \cite{Kobayashi:2011nu} to our specific 
situation, we obtain 
\be
\mathcal{G} ^{(2)} _{ab} = 
-\frac{1}{2} \, G_{2X} \nabla _a \phi \nabla _b \phi - \frac{1}{2}\, G_{2} 
\, g_{ab} \, ,
\ee

\be
\mathcal{G} ^{(3)} _{ab} =
\frac{1}{2} \, G_{3X} \Box \phi \nabla _a \phi \nabla _b \phi
+ \nabla_{(a} G_3 \nabla_{b)} \phi - \frac{1}{2} \, g_{ab} \nabla _c G_3 
\nabla ^c \phi \, ,
\ee

\be
\mathcal{G} ^{(4)} _{ab} = G_4 \, G_{ab} 
+ g_{ab} \left(G_{4\phi} \Box \phi - 2XG_{4\phi\phi} \right) 
- G_{4 \phi} \nabla _a \nabla _b \phi
-G_{4 \phi \phi} \nabla _a \phi \nabla _b \phi \,.
\ee

Recalling now that $\nabla _a G_3  = G_{3\phi} \nabla_a \phi + G_{3X} 
\nabla_a X$, one can write
\be
\begin{split}
\nabla_{(a} G_3 \nabla_{b)} \phi &= 
G_{3\phi} \nabla_{(a} \phi \nabla_{b)} \phi +G_{3X} \nabla_{(a} X 
\nabla_{b)} \phi\\
&= G_{3\phi} \nabla_{a} \phi \nabla_{b} \phi +G_{3X} \nabla_{(a} X 
\nabla_{b)} \phi 
\end{split}
\ee
and
\be
\begin{split}
\nabla _c G_3 \nabla ^c \phi &= G_{3\phi} \nabla_{c} \phi \nabla^{c} \phi 
+ G_{3X} \nabla_{c} X \nabla^{c} \phi\\
&= - 2 X G_{3\phi} + G_{3X} \nabla_{c} X \nabla^{c} \phi \,;\\
\end{split}
\ee
then we have that 
\be
\mathcal{G}^{(3)}_{ab} =
\frac{1}{2} \left( 2 G_{3\phi} + G_{3X} \Box \phi \right) \nabla _a \phi 
\nabla _b \phi
+G_{3X} \nabla_{(a} X \nabla_{b)} \phi
- \frac{1}{2}\, g_{ab} \left( - 2 X G_{3\phi} + G_{3X} \nabla X \cdot 
\nabla \phi \right) \, .
\ee
For future convenience we also rewrite $\mathcal{G}^{(4)}_{ab}$ as
\be
\mathcal{G} ^{(4)} _{ab} =
G_4 \, G_{ab} 
+ G_{4\phi} \left( g_{ab} \Box \phi - \nabla _a \nabla _b \phi \right) 
- G_{4 \phi \phi} \left( 2X g_{ab} + \nabla _a \phi \nabla _b \phi 
\right) \, .
\ee

We can now compute the effective stress-energy tensor of the $\phi$-fluid, 
which is defined by writing the  field equations of the specific sub-class 
of Horndeski theories as the effective 
Einstein equations
\be
G_{ab} = \frac{1}{G_4} \, T^{\rm (m)} _{ab} + T^{(\rm eff)} _{ab} \, , 
\label{effectiveEFE}
\ee
where the effective stress-energy tensor 
\be
T ^{(\rm eff)} _{ab} = T^{(2)} _{ab} + T^{(3)} _{ab} +T^{(4)} _{ab} 
\label{Teff}
\ee
with 
\be
T^{(2)} _{ab} = - \frac{\mathcal{G} ^{(2)} _{ab}}{G_4} = 
\frac{1}{2G_4} \left( G_{2X} \nabla _a \phi \nabla _b \phi + G_{2} \, 
g_{ab} \right) \, , \label{T2}
\ee
\be
\begin{split}
T^{(3)} _{ab} = - \frac{\mathcal{G} ^{(3)} _{ab}}{G_4} &= 
\frac{1}{2G_4}
\Big[ g_{ab} \left(G_{3X} \nabla X \cdot \nabla \phi  - 2 X G_{3\phi} 
\right) 
-2 G_{3X} \nabla_{(a} X \nabla_{b)} \phi \\
&\qquad \quad - \left( 2 G_{3\phi} + G_{3X} \Box \phi \right) \nabla _a 
\phi \nabla _b \phi \Big]
 \, , \label{T3}
\end{split}
\ee
\be
\begin{split}
T^{(4)} _{ab} = - \frac{\mathcal{G} ^{(4)} _{ab}- G_4 G_{ab}}{G_4}=
\frac{G_{4\phi}}{G_4} \left( \nabla _a \nabla _b \phi - g_{ab} \Box \phi 
\right)
+ \frac{G_{4\phi \phi}}{G_4}\left( 2X g_{ab} + \nabla _a \phi \nabla _b 
\phi \right)
\,. \label{T4}
\end{split}
\ee
The effective tensor given by Eqs.~(\ref{Teff})--(\ref{T4}) has the form of an 
imperfect fluid stress-energy tensor
\be
T_{ab} =  \rho u_a u_b + q_a u_b + q_b u_a + \Pi_{ab} 
\,, 
\ee
where
\be
\Pi_{ab} = T_{cd} {h_{a}}^c {h_{b}}^d= P  h_{ab} +\pi_{ab} 
\ee
is the effective stress tensor\footnote{The stress tensor $\Pi_{ab}$, the  
anisotropic stresses $\pi_{ab}$, and the heat current density  are purely 
spatial tensors,  $\Pi_{ab}u^a=\Pi_{ab}u^b=\pi_{ab}u^a=\pi_{ab}u^b=0$, 
$q_a u^a=0$.} containing the isotropic pressure  
\be
P = \frac{1}{3} g^{ab} \Pi _{ab} = \frac{1}{3} \, h^{ab} T _{ab} \, ,
\ee
the anisotropic stresses 
\be
\pi _{ab} = \Pi _{ab} - P \, h_{ab} \,,
\ee
the effective energy density
\be
\rho = T_{ab} u^a u^b \,,
\ee
and the effective heat flux density
\be
q_a = - T_{cd} u^c {h_{a}}^d \,. 
\ee
We compute these quantities separately for each contribution 
$T_{ab}^{(2,3,4)}$ to the effective energy-momentum tensor 
$ T^{(\rm eff)}_{ab} $. Let us begin with the  $T^{(2)}_{ab}$ 
contribution. Recalling that 
$\nabla _a \phi = \sqrt{2X} \, u_a$ and using the identities 
\begin{eqnarray}
&& h_{ab} u^b = h_{ab} u^a = 0 \,,\\
&&\nonumber\\
&& g_{ae} h^{eb} = {h_{a}}^b, \,\, {h^{a}}_e {h^{e}}_b 
={h^{a}}_b \,,\\
&&\nonumber\\
&& g_{ab} h^{ab} = {h^{a}}_a = 3 \,,
\end{eqnarray}
one finds
\be
T^{(2)} _{ab} =  
\frac{1}{2G_4} \left( 2G_{2X} X u_a u_b + G_{2} \, g_{ab} \right) 
\ee
and the effective fluid quantities for this part of the 
effective stress-energy tensor of the $\phi$-fluid are
\be
\begin{split}
\rho ^{(2)} = T^{(2)}_{ab} u^a u^b &= \frac{1}{2G_4} \left( G_{2X} 2X u_a 
u_b + G_{2} \, g_{ab} \right) u^a u^b \\
&=\frac{1}{2G_4} \left(2X G_{2X} - G_{2} \right) \,,
\end{split}
\ee

\be
\begin{split}
q_a ^{(2)} = - T^{(2)}_{cd} u^c {h_{a}}^d &= 
- \frac{1}{2G_4} \left( G_{2X} 2X u_c u_d + G_{2} \, g_{cd} \right) u^c 
{h_{a}}^d \\
&=- \frac{1}{2G_4} \left(- 2X G_{2X} u_d + G_{2} \, u_{d} \right) 
{h_{a}}^d = 0 \,,  
\end{split}
\ee

\be
\begin{split}
\Pi ^{(2)} _{ab} = T ^{(2)} _{cd} {h_{a}}^c {h_{b}}^d &= 
\frac{1}{2G_4} \left( G_{2X} 2X u_c u_d + G_{2} \, g_{cd} \right) 
{h_{a}}^c {h_{b}}^d \\
&= \frac{G_2}{2G_4} \, h_{ab} \, , 
\end{split}
\ee

\be
\begin{split}
P ^{(2)} =\frac{1}{3} \, g^{ab} \Pi^{(2)} _{ab} &= \frac{1}{3} \, g^{ab} 
\, \frac{G_2}{2G_4} \, h_{ab} 
= \frac{1}{3}\, {h^{a}}_a \, \frac{G_2}{2G_4} = \frac{G_2}{2G_4} \,, 
\end{split}
\ee

\be
\begin{split}
\pi ^{(2)} _{ab} = \Pi^{(2)} _{ab} - P^{(2)} \, h_{ab} = 0 \, .
\end{split}
\ee

Continuing with the effective fluid quantites associated with the 
contribution  $T^{(3)}_{ab}$ to $ T^{(\rm eff)}_{ab} $, one finds
\begin{equation}
\begin{split}
T^{(3)} _{ab} = \frac{1}{2G_4}
\Big[ g_{ab} \left(G_{3X} \nabla X \cdot \nabla \phi  - 2 X G_{3\phi} \right) 
-2 \sqrt{2X} G_{3X} \nabla_{(a} X u_{b)}
- 2X \left( 2 G_{3\phi} + G_{3X} \Box \phi \right) u _a u _b \Big]
 \, ,
\end{split}
\end{equation}
and

\be
\begin{split}
\rho ^{(3)} &= T^{(3)}_{ab} u^a u^b\\
&= \frac{1}{2G_4} 
\Big[ - \left(G_{3X} \nabla X \cdot \nabla \phi  - 2 X G_{3\phi} \right) 
+2 G_{3X} \sqrt{2X} u^a \nabla_{a} X \\
&\qquad \qquad- 2X \left( 2 G_{3\phi} + G_{3X} \Box \phi \right) \Big]\\
&= - \frac{1}{2G_4} \left( - G_{3X} \nabla X \cdot \nabla \phi + 2X 
G_{3\phi} + 2X G_{3X} \Box \phi \right) \, .
\end{split}
\ee

To compute $q_a ^{(3)}$ one makes use of the facts that
\begin{eqnarray}
&& g_{cd} \, u^c {h_a}^d =u_d \, {h_a}^d = 0 \, ,\\
&&\nonumber\\
&& u^c \, {h_a}^d  \, \nabla_{(a} X u_{b)} = 
\frac{1}{2} \, u^c \, {h_a}^d \left( \nabla_{a} X u_{b} + \nabla_{b} X 
u_{a} 
\right) = - \frac{1}{2} \, h_{ad} \nabla^d X \,,\\
&&\nonumber\\
&& u_c u_d u^c {h_a}^d = 0 \,,
\end{eqnarray}
that yield
\be
\begin{split}
q_a ^{(3)} &= - T^{(3)}_{cd} u^c {h_{a}}^d 
= - \frac{G_{3X}}{2 G_4} \, \sqrt{2X} \left(\nabla_a X + \frac{\nabla X 
\cdot \nabla \phi}{2X} \, \nabla_a \phi \right) \, .
\end{split}
\ee
Now, to compute $\Pi ^{(3)} _{ab}$ one uses
\begin{eqnarray}
&& {h_a}^c \, {h_b}^d \, g_{cd} = h_{ab} \, ,\\
&&\nonumber\\
&& {h_a}^c \, {h_b}^d \, \nabla_{(a} X u_{b)} =
\frac{1}{2} \, {h_a}^c \, {h_b}^d \, \left( \nabla_{a} X u_{b} + 
\nabla_{b} X u_{a} \right) = 0 \,,\\
&&\nonumber\\
&&  {h_a}^c \, {h_b}^d \, u_c u_d = 0 \, ,
\end{eqnarray}
which lead to
\begin{eqnarray}
&& \Pi ^{(3)} _{ab} = \frac{h_{ab}}{2 G_4} \left(G_{3X} \nabla X \cdot 
\nabla \phi  - 2 X G_{3\phi} \right) \, ,\\
&&\nonumber\\
&& P^{(3)} =\frac{1}{3} g^{ab} \Pi^{(3)} _{ab}
=  \frac{1}{2 G_4} \left(G_{3X} \nabla X \cdot \nabla \phi  - 2 X 
G_{3\phi} \right) \, ,
\end{eqnarray}
and
\be
\pi ^{(3)} _{ab} = \Pi^{(3)} _{ab} - P^{(3)} \, h_{ab} = 0 \, .
\ee

Finally, the $T^{(4)}_{ab}$ contribution 
\be
T^{(4)} _{ab} = 
\frac{G_{4\phi}}{G_4} \left( \nabla _a \nabla _b \phi - g_{ab} \Box \phi 
\right) + 2 X \,\frac{G_{4\phi \phi}}{G_4}
\left( g_{ab} + u_a \, u_b \right)\, ,
\ee
is calculated using the intermediate result reported in 
Appendix~\ref{sec:AppendixA}, obtaining
\be
\begin{split}
\rho ^{(4)} &= T^{(4)}_{ab} u^a u^b = u^a u^b \left[ 
\frac{G_{4\phi}}{G_4} \left( \nabla _a \nabla _b \phi - g_{ab} \Box \phi 
\right) 
+ 2 X \,
\frac{G_{4\phi \phi}}{G_4}
\left( g_{ab} + u_a \, u_b \right) \right]\\
&= \frac{G_{4\phi}}{G_4} \left( \Box \phi - \frac{\nabla X \cdot \nabla 
\phi}{2X} \right)\, , 
\end{split}
\ee

\be
\begin{split}
q_a ^{(4)} &= - T^{(4)}_{cd} u^c {h_{a}}^d\\
&= - \frac{G_{4\phi}}{G_4} \, u^c {h_{a}}^d \, \nabla _c \nabla _d \phi\\
&= - \frac{G_{4\phi}}{G_4} \, \frac{\nabla^c \phi}{\sqrt{2X}} 
\left(\nabla _c \nabla _a \phi + \frac{\nabla_a \phi \nabla^d \phi \nabla 
_c \nabla _d \phi}{2X} \right) \\
&= \frac{G_{4\phi}}{G_4} \, \frac{1}{\sqrt{2X}} 
\left(\nabla _a X + \frac{\nabla \phi \cdot \nabla X}{2X} \, \nabla_a 
\phi \right)
\end{split}
\ee
and 
\be
\begin{split}
\Pi ^{(4)} _{ab} &= T ^{(4)} _{cd} {h_{a}}^c {h_{b}}^d \\
&= \left[ \frac{G_{4\phi}}{G_4} \left( \nabla _c \nabla _d \phi - g_{cd} 
\Box \phi \right) 
+ 2 X \, \frac{G_{4\phi \phi}}{G_4}\, g_{cd} + (\propto u_c \, u_d ) 
\right] \, {h_{a}}^c {h_{b}}^d \\
&= \frac{G_{4\phi}}{G_4} \, {h_{a}}^c {h_{b}}^d \nabla _c \nabla _d \phi
+ h_{ab} \left( 2 X \, \frac{G_{4\phi \phi}}{G_4} - \frac{G_{4\phi}}{G_4} 
\, \Box \phi \right) \,,
\end{split}
\ee
or
\be
\begin{split}
\Pi ^{(4)} _{ab} &=  
\frac{G_{4\phi}}{G_4} 
\left( \nabla _a \nabla _b \phi 
- \frac{\nabla_{(a} X \nabla_{b)} \phi}{X}
- \frac{\nabla X \cdot \nabla \phi}{4X^2} \, \nabla_a \phi \nabla_b \phi 
\right)
+ h_{ab} \left( 2 X \, \frac{G_{4\phi \phi}}{G_4} - \frac{G_{4\phi}}{G_4} 
\, \Box \phi \right) \, ,
\end{split}
\ee
while

\be
\begin{split}
P ^{(4)} &=\frac{1}{3} g^{ab} \Pi^{(4)} _{ab} \\
&= - \frac{G_{4\phi}}{3 G_4} \left( 2 \Box \phi + \frac{\nabla X \cdot 
\nabla \phi}{2X} \right)
+ 2X \, \frac{G_{4\phi \phi}}{G_4} \,,
\end{split}
\ee

\be
\begin{split}
\pi ^{(4)} _{ab} &= \Pi^{(4)} _{ab} - P^{(4)} \, h_{ab} =\\
&=\frac{G_{4\phi}}{G_4} \left[ \nabla _a \nabla _b \phi 
- \frac{\nabla_{(a} X \nabla_{b)} \phi}{X}
- \frac{\nabla X \cdot \nabla \phi}{4X^2} \, \nabla_a \phi \nabla_b \phi
+ \frac{h_{ab}}{3} \left( \frac{\nabla X \cdot \nabla \phi}{2X} - \Box 
\phi \right)\right] \,.
\end{split}
\ee
To summarize, the field equations of the chosen subclass of Horndeski 
theories of gravity have been rewritten in the form of effective Einstein 
equations by moving the Horndeski terms to their right hand side and 
leaving the Einstein tensor in the left hand side. It is a fact that the right 
hand side of the field equations, recast in this form, assumes the form 
of the stress-energy tensor of a {\em dissipative} effective fluid. 
Thus far, only a manipulation of the  field equations of the specific subclass of 
Horndeski theories has been 
performed and no extra assumption has been made. The results 
presented in this Section confirm the ones discussed 
in~\cite{Quiros:2019gai}.

\section{Thermodynamic analogy for the effective dissipative $\phi$-fluid}
\label{sec:5}
\setcounter{equation}{0}

We are now ready to examine the consequences of writing the Horndeski 
field equations (for the class of Horndeski theories considered) in the 
form of Einstein equations with an effective dissipative fluid. Although 
this reduction has been performed many times in the literature in various 
special contexts (including Brans-Dicke or $f(R)$ gravity, nonminimally 
coupled scalar fields, Friedmann-Lema\^itre-Robertson-Walker metrics or 
cosmological perturbations in extended gravity), the physical 
interpretation of the effective dissipative fluid and of its 
thermodynamics is usually not attempted. We began looking for this 
physical intepretation, for simple Brans-Dicke-like and $f(R)$ gravity, in 
\cite{Faraoni:2021lfc,Mentrelli}. To this end, we adopt the most basic 
elements of Eckart's theory of gravity \cite{Eckart:1940te}. While it is well 
known that this theory is riddled with causality violation and 
instabilities, it is nevertheless the model of dissipative fluid most 
frequently used in relativity \cite{Maartens:1996vi,Andersson:2006nr}. We 
assume the constitutive equations of 
Eckart's theory: these are phenomenological equations that could be 
assumed in a variety of theories of dissipation and constitute minimal 
assumptions on the physics of a (real or effective) dissipative fluid 
\cite{Eckart:1940te,Maartens:1996vi,Andersson:2006nr}.

The three constitutive equations (\cite{Eckart:1940te}, 
see also \cite{Andersson:2006nr}), relate the viscous pressure $P_\text{vis}$ 
with the fluid expansion $\theta$, the heat current density $q^a$
with  the temperature ${\cal T}$, and the anisotropic stresses $\pi_{ab}$ with 
the  shear tensor $\sigma_{ab}$:
\begin{eqnarray}
P_\text{vis} &=& -\zeta \, \theta \,,\label{def:zeta}\\
&&\nonumber\\
q_a &=& -{\cal K} \left( h_{ab} \nabla^b {\cal T} + {\cal T} \dot{u}_a 
\right)
\,, \label{Eckart}\\
&&\nonumber\\
\pi_{ab} &=& - 2\eta \, \sigma_{ab} \,,\label{def:eta}
\end{eqnarray}
where $\zeta$, ${\cal K}$, and $\eta$ are the thermal conductivity, bulk viscosity, and
shear viscosity, respectively.

Let us begin with the phenomenological extension of Fourier's law relating  
heat flux density and temperature. The calculations of the previous 
section provide the effective heat flux density
\be
\label{eqqeff}
q_a ^{(\rm eff)} = q_a ^{(3)} + q_a ^{(4)} = \frac{G_{4\phi} - X 
G_{3X}}{G_{4} \sqrt{2X}}  \left( \nabla _a X + \frac{\nabla \phi \cdot 
\nabla X}{2 X} \nabla _a \phi \right) 
\ee
in the subclass of Horndeski theories. One infers from Eq. 
\eqref{eq:udot} that
\be
q_a ^{(\rm eff)} = 
- \frac{\sqrt{2 X} (G_{4 \phi} - X G_{3 X})}{G_{4}} \, \dot{u}_a \,.
\ee
Turning to Eq.~(\ref{Eckart}), it turns out \cite{Faraoni:2018qdr, 
Faraoni:2021lfc,Mentrelli} that for ``old'' scalar-tensor 
gravity, the spatial 
gradient  $h_{ab} \nabla^b \mathcal{T}$ vanishes in the comoving 
frame,\footnote{The spatial temperature gradient of a fluid does not always 
vanish in the frame comoving with it: for example, in a static fluid in thermal 
equilibrium in a 
static gravitational field, the temperature obeys the Tolman condition ${\cal 
T}\sqrt{-g_{00}}=$~const. \cite{MTW}.}
 leaving only the inertial term in the heat flux density \be
\label{eqComovingEckart}
q_a = - \mathcal{K} \mathcal{T} \, \dot{u} _a  \, .
\ee

Comparing Eq. \eqref{eqqeff} and \eqref{eqComovingEckart}, one can make 
the identifications 
\be
\label{E-1}
{\cal K} \, h_{ab} \nabla^b {\cal T} = 0 \, ,
\ee
and
\be
\label{E-2}
\mathcal{K} \mathcal{T} \equiv \frac{\sqrt{2 X} (G_{4 \phi} - X G_{3 
X})}{G_{4}}  \, ,
\ee
where $\mathcal{K}$ and ${\cal T}$ denote, respectively, the thermal conductivity and 
effective temperature of the $\phi$-fluid for the subclass Horndeski gravity. 
Here ${\cal T}$ is the ``temperature of gravity'', which reduces to the 
quantity already identified in ``old'' ({\em i.e.}, Brans-Dicke-like) 
scalar-tensor theories in 
Refs.~\cite{Faraoni:2018qdr,Faraoni:2021lfc,Mentrelli}.

To continue on the lines of \cite{Faraoni:2021lfc,Mentrelli}, we identify 
a shear viscosity for the effective $\phi$-fluid. The latter has 
anisotropic stress tensor 
\be
\label{eqstress}
\pi_{ab} ^{(\rm eff)} =  \pi ^{(4)} _{ab} =
\frac{G_{4 \phi}}{G_{4}} \left[ \nabla _a \nabla_b \phi - 
\frac{\nabla_{(a} X \nabla _{b)} \phi}{X} 
 - \frac{\nabla X \cdot \nabla \phi}{4 X^2} \nabla _a \phi \nabla _b \phi
- \frac{h_{ab}}{3} \left( \Box \phi - \frac{\nabla X \cdot \nabla \phi}{2 
X} \right) \right] 
\ee
and, from Eq. \eqref{eq:shear}, one infers that
\be
\pi_{ab} ^{(\rm eff)} = \frac{G_{4\phi} \sqrt{2X}}{G_{4}} \,\sigma _{ab} 
\,.
\ee
We now assume the second constitutive equation of Eckart's 
theory relating the anisotropic stresses $\pi_{ab}$ with the shear tensor 
$\sigma_{ab}$ in a dissipative fluid \cite{Eckart:1940te}
\be
\label{eqEckartShear}
\pi_{ab} = - 2 \eta \, \sigma _{ab} \,,
\ee
where $\eta$ is the shear viscosity. Comparing Eqs. \eqref{eqqeff} and 
\eqref{eqComovingEckart}, one is naturally led to identify
\be
\eta = - \frac{\sqrt{X} \, G_{4 \phi}}{\sqrt{2} \, G_{4}} 
\ee
with the shear viscosity of the effective $\phi$-fluid, where it is 
$G_4>0$ to guarantee a positive gravitational coupling of gravity to 
matter. Since one can always redefine the scalar field $\phi$ according to 
$\psi= G_4(\phi)$ (this relation is invertible whenever $G_{4 \phi}\neq 
0$), the shear viscosity becomes $ \eta = - \frac{\sqrt{X} \, \psi_{, 
\phi} }{\sqrt{2} \, \psi} $ and is positive whenever $G_{4\phi}<0$ and 
negative otherwise, for example in Brans-Dicke theory where 
$G_4(\phi)=\phi$ \cite{Brans:1961sx}. Negative viscosities can occur 
in fluid mechanics, atmospheric physics, ocean currents, liquid 
crystals, {\em etc.} Typically, they are related with turbulence and occur 
in non-isolated systems which receive energy from the outside (see, {\em 
e.g.}, Refs.~\cite{negviscosity}). Indeed, the effective $\phi$-fluid is 
not isolated since the scalar $\phi$ couples explicitly to gravity through 
the term $G_4  R $ in the Horndeski Lagrangian.

The structure of  $T^{(\rm eff)}_{ab}$ (in the form that we have chosen) 
does not allow for a viscous pressure, hence the bulk viscosity vanishes,  
$\zeta=0$. 
It is interesting 
that, in the very different context of spacetime thermodynamics in 
$f({\cal R})$ gravity, bulk viscosity is absent and shear viscosity is 
important (this fact was emphasized in Ref.~\cite{Chirco:2010sw} and 
corrects the previous interpretation of \cite{Eling:2006aw} of the 
thermodynamics of spacetime in $f(R)$ gravity).

A general interpretation of $\mathcal{K}$ and $\mathcal{T}$ emerges from 
the thermodynamic analogy. If one chooses
\be 
\label{K-solution}
\mathcal{K} \equiv \sqrt{2X} \left( G_{4\phi} - X \, 
G_{3X}\right) 
\ee
and
\be 
\label{T-solution}
\mathcal{T} \equiv \frac{1}{G_4} \, ,
\ee
then $\mathcal{T}$ automatically satisfies $h_{ab} \nabla^b \mathcal{T} 
=0$. Indeed, $\mathcal{T} = \mathcal{T} (\phi)$ since $G_4 =G_4 (\phi)$, 
thus $\nabla_a \mathcal{T} \propto \nabla_a \phi$. Furthermore, it must be 
$G_4>0$ to guarantee a positive gravitational coupling strength of gravity 
to matter, as is clear from Eq.~(\ref{effectiveEFE}), and the 
``temperature'' of gravity ${\cal T}$ is non-negative. This fact was not 
guaranteed {\em a priori}. GR corresponds to $\phi=$~const. and, 
therefore, 
to a unit value of the temperature (if coupling with matter is considered) 
and vanishing thermal conductivity  
in the spectrum of the specific subclass of Horndeski theories. 
 This fact embodies the idea that GR is a 
``state of equilibrium'' in a wider space of theories of gravity and 
any extension of gravity corresponds to a deviation from equilibrium, which is 
rather natural if extra degrees of freedom (in this case the scalar 
$\phi$) are excited.

Lastly, it is worth noting that if one expands the general Horndeski action on a 
spatially flat, homogeneous, and isotropic background to second order in the
linear perturbations, then the corresponding dynamics is controlled by 
four functions of time (see {\em e.g.} \cite{Bellini:2014fua,Noller}, and references therein). 
Among these functions one finds the effective Planck mass $M_{\star} ^2$ and
the braiding $\alpha _{\rm B}$, quantifying the strength of the kinetic mixing 
between scalar and tensor perturbations, that for the case of the specific subclass 
of Horndeski theories simply read
\be
\label{parameters}
M_{\star} ^2 = 2 \, G_4 
\qquad \mbox{and} \qquad 
\alpha _{\rm B} = \frac{2 \, \dot{\phi}}{H M^2 _\star} \left( X \, 
G_{3X} - G_{4\phi} \right)  \, ,
\ee    
where $\phi$, $G_3$, $G_4$, and their derivatives are evaluated on the
background configuration and with $H$ denoting the Hubble parameter.
A comparison between \eqref{K-solution}, \eqref{T-solution}, and \eqref{parameters}
shows that $\mathcal{T}\propto1/M_{\star} ^2$ and $\mathcal{K} \propto -\alpha _{\rm B}$ 
in this specific realization of the analogy. This suggests a deeper physical significance 
behind the choice of \eqref{K-solution} and \eqref{T-solution}, beyond their simplicity, 
among the broader class of solutions of the system \eqref{E-1}-\eqref{E-2}.
However, it is worth stressing that the derivation of the functions $M_{\star} ^2$
and $\alpha _{\rm B}$ is performed on cosmological backgrounds, which is a context that
still escapes the formalism discussed here \cite{Faraoni:2021lfc, Mentrelli}. 
Hence, any further consideration on this matter is postponed to future investigations.

\section{More general Horndeski theories}
\label{sec:6}
\setcounter{equation}{0}

We now move on to discuss more general Horndeski theories of gravity. 
Consider an Horndeski model such that 
\be
\delta \left( \sqrt{-g} \, \mathcal{L} \right) = \sqrt{-g} \, \mathcal{G} 
_{ab} \, \delta g^{ab} 
+ (\cdots) \, \delta \phi + \mbox{total derivative} \, ,
\ee
with
\be
\label{bruttaroba}
\mathcal{G}_{ab} \supset \xi(\phi, X) \, R_{a c b d} \nabla^c \phi 
\nabla^d \phi \, ,
\ee
which is a common feature of theories beyond the subclass we considered so far 
(see Ref.~\cite{Kobayashi:2011nu} for the corresponding field 
equations). As it becomes clear, they contain derivative non-minimal couplings. This choice implies
\be
T_{ab} \supset \zeta(\phi, X) \, R_{a c b d} \nabla^c \phi \nabla^d \phi \, ,
\ee
where $\zeta(\phi, X)$ is a function proportional to $\xi(\phi, X)$; from this  
one concludes that
\be
\Pi _{ab} = T _{cd} {h_{a}}^c {h_{b}}^d \supset 
\zeta(\phi, X) \, {h_{a}}^c  {h_{b}}^d  R_{c e d f} \nabla^e \phi \nabla^f 
\phi =
\zeta(\phi, X) \, R_{a e b f} \nabla^e \phi \nabla^f \phi \, ,
\ee
taking advantage of the symmetries of the Riemann tensor, which lead to 
$ R_{c e d f}  \nabla^e \phi \nabla^f \phi \nabla^c \phi 
\nabla^d \phi = 0 $. Taking the trace of the stress tensor yields 
\be
P =\frac{1}{3} g^{ab} \Pi _{ab} \supset \frac{\zeta(\phi, X)}{3} \, g^{ab} \, 
R_{a e b f} \nabla^e \phi \nabla^f \phi =
\frac{\zeta(\phi, X)}{3} \, R_{e f} \nabla^e \phi \nabla^f \phi 
\ee
and then we have
\be
\pi _{ab} = \Pi _{ab} - P \, h_{ab} \supset 
\zeta(\phi, X) \, R_{a e b f} \nabla^e \phi \nabla^f \phi - 
\frac{\zeta(\phi, X)}{3} \, R_{e f} \nabla^e \phi \nabla^f \phi \, .
\ee
While the term containing the Ricci tensor can, in principle, cancel out 
with similar terms coming from the field equations and contained in the 
effective energy-momentum tensor, the contribution to the right hand side  
depending on the Riemann tensor cannot be traced away.\footnote{Although 
it is not {\em a priori} unconceivable that this contribution is  
removed by imposing some relation between $G_4$ and 
$G_5$, the latter would be extremely fine-tuned and would give rise to 
a completely artificial theory.} This Riemann tensor term 
breaks the 
proportionality between $\sigma _{ab}$ (which is a kinematic quantity and, 
therefore, does not depend on the specific model) and $\pi _{ab}$. As a 
consequence, the three constitutive equations of Eckart's theory 
\cite{Eckart:1940te}  no longer 
hold for the effective fluid.  Of course, this 
proportionality could be 
broken by other terms coming from the variation of $\mathcal{L}_4$ and 
$\mathcal{L}_5$, though the feature discussed here
involves a property that seem to be shared by the vast majority of 
models beyond the specific luminal Horndeski class.

As an example,  let us discuss the Horndeski theory, not belonging to the 
restricted class \eqref{eq:LsH}, with Lagrangian density
\be
\mathcal{L}=G_4(X) R =X R\,,
\ee
for which the effective energy-momentum tensor of the $\phi$-fluid is
\begin{eqnarray}
T_{ab}^{(\rm eff)} &=& \frac{1}{X} \left\{ 
\frac{R}{2} \, \nabla_a\phi \nabla_b \phi +\Box \phi \nabla_a\nabla_b \phi 
- \nabla_a \nabla_e \phi \nabla^e \nabla_b \phi
-\frac{1}{2} \, g_{ab} \left[ \left( \Box \phi \right)^2 -\left( 
\nabla\nabla \phi \right)^2\right] 
-2 R_{e(a}\nabla_{b)} \phi \nabla^e \phi \right.\nonumber\\
&&\nonumber\\
&\, & \left. +g_{ab} R^{ef} \nabla_e \phi 
\nabla_f \phi -R_{aebf} \nabla^e\phi \nabla^f\phi \right\} \,,
\end{eqnarray}
where $
\left( \nabla\nabla \phi \right)^2 \equiv \nabla^c\nabla^d\phi 
\nabla_c \nabla_d\phi$. The effective stress tensor is then
\begin{eqnarray}
\Pi_{ab}^{(\rm eff)} &=& T^{(\rm eff)}_{cd} {h_a}^c {h_b}^d \nonumber\\
&&\nonumber\\
&=& \frac{1}{X} \left\{ \Box \phi \nabla_c\nabla_d\phi 
-\nabla_c\nabla_e \phi \nabla^e \nabla_d \phi 
-\frac{1}{2} \, g_{cd}\left[  \left( \Box \phi \right)^2 -\left( 
\nabla\nabla \phi \right)^2\right] +g_{cd} R^{ef} \nabla_e\phi\nabla_f\phi 
-R_{cedf} \nabla^e\phi \nabla^f\phi \right\} {h_a}^c {h_b}^d 
\,.\nonumber\\
&& \label{chimelafattofare}
\end{eqnarray}
The terms ${h_a}^c {h_b}^d \Box\phi 
\nabla_c\nabla_d\phi$ and 
$ {h_a}^c {h_b}^d \nabla^e\nabla_c \phi \nabla_e\nabla_d\phi$ appearing in 
this expression are computed in Appendix~\ref{sec:AppendixB}; substituting 
them into Eq.~(\ref{chimelafattofare}), one obtains
\begin{eqnarray}
\Pi_{ab}^{(\rm eff)} &=& \frac{1}{X} \left\{ \Box\phi \left[ 
\nabla_a\nabla_b \phi -\frac{1}{X} \, \nabla_{(a}X \nabla_{b)}\phi 
-\frac{1}{4X^2} \, \nabla_a \phi \nabla_b \phi \nabla^c\phi \nabla_c 
X\right] - \nabla_a\nabla_e \phi \nabla^e \nabla_b \phi +\frac{1}{X} \, 
\nabla_{(a }\phi \nabla_{b)} \nabla_e \phi \nabla^e X \right. \nonumber\\
&&\nonumber\\
&\,&\left.   - \nabla_a\phi 
\nabla_b \phi \, \frac{ \nabla_e X \nabla^e X}{4X^2} 
-\frac{1}{2} \, h_{ab} \left[ \left( \Box\phi \right)^2 -\left( \nabla  
\nabla\phi\right)^2 \right] 
+h_{ab} R^{ef} \nabla_e\phi \nabla_f\phi  - R_{cdef} {h_a}^c {h_b}^d  
\nabla^e\phi \nabla^f \phi \right\} \,.
\end{eqnarray}
From this one computes the effective fluid pressure
\begin{eqnarray}
P^{(\rm eff)} &=& \frac{1}{3} \, g^{ab} \Pi_{ab}^{(\rm eff)} \nonumber\\
&&\nonumber\\
&=& \frac{1}{3X} \left[ -\frac{ \left(\Box \phi\right)^2}{2} +\frac{3}{2} 
\, 
\left( \nabla\nabla\phi \right)^2 -\frac{\Box\phi}{2X} \, 
\nabla^c\phi\nabla_c X -\frac{\nabla^c X \nabla_c X}{2X} +3R_{ef} 
\nabla^e\phi \nabla^f \phi - \nabla^a\nabla^e\phi \nabla_a\nabla_e \phi 
\right] \,.
\end{eqnarray} 

The anisotropic stresses are then obtained as
\begin{eqnarray}
\pi_{ab}^{(\rm eff)} & \equiv & \Pi_{ab}^{(\rm eff)} -P^{(\rm eff)} h_{ab} 
\nonumber\\
&&\nonumber\\
&=& \frac{1}{X} \left\{ \Box \phi \left[  
\nabla_a\nabla_b\phi -\frac{\nabla_{(a}\phi \nabla_{b)}X}{X}
-\frac{ \nabla^c \phi \nabla_c X}{4X^2} \, \nabla_a\phi \nabla_b \phi 
+\frac{ \nabla^c \phi \nabla_c X}{6X} \, h_{ab} \right] \right.\nonumber\\
&&\nonumber\\
&\, &\left. - \nabla_a \nabla_e \phi \nabla^e\nabla_b\phi +
\frac{ \nabla_{(a}\phi \nabla_{b)}\nabla_e \phi \nabla^e X}{X} - 
\nabla_a\phi 
\nabla_b\phi \, \frac{ \nabla^e X \nabla_e X}{4X^2} +
\frac{ \nabla^e X \nabla_e X}{6X} \, h_{ab} -\frac{ \left( \Box 
\phi\right)^2}{6} \, h_{ab} \right.\nonumber\\
&&\nonumber\\
&\, & \left.  - R_{cdef} {h_a}^c {h_b}^d \nabla^e \phi \nabla^f \phi 
+\frac{1}{3} \, 
\left( \nabla^a \nabla^e \phi \nabla_a \nabla_e \phi \right) 
h_{ab}\right\} \,.
\end{eqnarray} 
This quantity is definitely not proportional to the shear 
$\sigma_{ab}^{(\rm eff)}$, thus breaking the analogy with 
Eckart's constitutive relation~(\ref{def:eta}). Similarly, one computes 
the effective heat current density
\be
q_a^{({\rm eff})} = -\frac{1}{X} \left[ \Box \phi \, \nabla_c\nabla_d \phi 
-\nabla_c \nabla_e \phi \nabla^e \nabla_d \phi -R_{cedf} \nabla^e\phi \nabla^f\phi \right] u^c {h_a}^d \,,
\ee
which cannot be reduced to the Eckart constitutive 
relation~(\ref{Eckart}). 

The presence of non-minimal derivative couplings leads to terms in the ﬁeld equations of the form of \eqref{bruttaroba} that break the thermodynamic description. Most notably, these terms are those operators which quite generically forbid a dual description in the Jordan frame due to intrinsic changes in the gravity sector. In terms of a non-local ﬁeld redeﬁnition, one could write the dual description in the ``Jordan'' frame where the eﬀective ﬂuid and its stress-energy tensor would become non-local. However, in terms of a local description the separation between ``gravity'' and ``matter fluid'' fails apart and the same happens to the analogy with Eckart's theory.

\section{Discussion and conclusions}
\label{sec:7}
\setcounter{equation}{0}

For first generation of scalar-tensor theories, the description in terms of 
an effective $\phi$-fluid which is dissipative leads, through Eckart's 
first order thermodynamics \cite{Eckart:1940te}, to a formalism of 
``thermodynamics of gravity'' in which GR is seen as the state of 
equilibrium and scalar-tensor gravity, with its extra scalar degree of 
freedom $\phi$, as a non-equilibrium state. In many situations, the 
dissipation leads to an approach to GR  
\cite{Faraoni:2021lfc,Mentrelli}. It is natural to apply a similar 
formalism to Horndeski theories of gravity, which generalize ``old'' 
scalar-tensor gravity \cite{Brans:1961sx,Bergmann:1968ve, 
Nordtvedt:1968qs, Wagoner:1970vr,Nordtvedt:1970uv} and which have seen an 
explosion of activity during the past decade.

The first step consists of  extracting an effective dissipative fluid of 
the scalar 
degree of freedom $\phi$ from the field equations. This step was started in 
Ref.~\cite{Quiros:2019gai}, which we generalize, introducing minor 
corrections. In 
particular, the kinematic quanties $u^a$, $\dot{u}^a, \theta, 
\omega_{ab}, V_{ab}$, 
and $\sigma_{ab}$ are identical to the previous derivations 
in\footnote{The previous Ref.~\cite{Pimentel89} did not provide the 
kinematical quantities.}  Refs.~\cite{Faraoni:2018qdr,Faraoni:2021lfc}, 
since they do not depend on the field equations.  We 
emphasize that the energy-momentum tensor of the effective $\phi$-fluid 
obtained does not satisfy any energy condition, nor it is expected to: it 
is built out of gravitational terms and does not arise from a kinetic 
theory. In spite of this shortcoming from a fluid-mechanicist's  
point of view, its interpretation as a 
dissipative fluid \`a la Eckart can provide an intriguing view of GR as 
the (constant) unit temperature state, with vanishing thermal 
conductivity, 
and of Horndeski gravity as a non-equilibrium state. 
This view is independent of Jacobson's thermodynamics of spacetime 
\cite{Jacobson:1995ab,Eling:2006aw}, but it echoes two of its main ideas.

The next step consists of applying Eckart's thermodynamics to this 
effective fluid. In comparison with Jacobson's thermodynamics of spacetime 
\cite{Jacobson:1995ab,Eling:2006aw}, the effective fluid approach and 
Eckart description are minimalist in their assumptions.

The calculations are a bit tedious, according to how many terms are 
allowed in the Horndeski action. It turns out that our approach does not 
work for the most general Horndeski theory: even though one can define the 
effective fluid, including its heat current density and anisotropic 
stresses, they do not satisfy the constitutive relations 
(\ref{def:zeta})--(\ref{def:eta}) linking them with the viscous pressure, 
shear, fluid four-acceleration, and temperature gradient in Eckart's 
formalism (or in any thermodynamical theory in which reasonable 
constitutive relations are needed). However, when the terms violating the 
constraints on the speed of gravitational waves are dropped from the 
Horndeski action, the ``temperature of gravity'' formalism makes sense 
again, the temperature ${\cal T}$ is positive-definite, and GR 
corresponds, for instance, to ${\cal T} = 1$ and ${\cal K} = 0$. Although, 
this interpretation is not the only one possible since ${\cal T}$ and 
${\cal K}$ are ultimately defined by the system of equations given by 
\eqref{E-1} and \eqref{E-2}.  One could take this result to say  
that there could be a relation between physical constraints such as those 
usually imposed on Horndeski gravity (related to stability and the 
propagation of gravitational waves) and the validity of the thermodynamic 
analogy relating kinematic quantities and the components of the 
$\phi$-fluid through Eckart's constitutive equations.  The effective 
temperature of gravity formalism is still under development and 
far-fetching conclusions are premature, however this result is rather 
suggestive.

In any case, although intriguing, the approach followed here suffers from 
the limitations intrinsic to Eckart's first-order thermodynamics 
\cite{Eckart:1940te} (or better, of its constitutive equations, which is 
all that was used here). An attempt to generalize the present work to 
causal (second-order) thermodynamics will be presented elsewhere.
Similarly, one can classify different extensions of GR based on their ``thermodynamic running"
to the GR fixed point, specifically based on the presence or absence of hairy solutions away from GR.
Further, an investigation of the effective fluid description of vector-tensor theories 
\cite{Heisenberg:2014rta}, and potential connections with non-equilibrium thermodynamics, 
is left to a future study.

\appendix
\section{Useful relations}
\label{sec:AppendixA}
\renewcommand{\theequation}{A.\arabic{equation}}
\setcounter{equation}{0}

The following relations are useful to compute the various contributions 
to the stress-energy tensor of the effective dissipative fluid 
associated with the scalar field $\phi$:
 \be 
\nabla \phi \cdot \nabla X = - \nabla ^a \phi 
(\nabla _c \phi \nabla _a \nabla ^c\phi) = 
- \nabla^a \phi \nabla^b \phi \nabla _a \nabla _b \phi \,,
\ee

\be
\begin{split}
{h_{b}}^d \nabla _c \nabla _d \phi &= 
\left({\delta_{b}}^d + \frac{\nabla_b \phi \nabla^d \phi}{2X} \right) 
\nabla _c \nabla _d \phi
= \nabla _c \nabla _b \phi + \frac{\nabla_b \phi \nabla^d \phi \nabla _c 
\nabla _d \phi}{2X} \,,
\end{split}
\ee
and
\be
\begin{split}
{h_{a}}^c \,{h_{b}}^d \nabla _c \nabla _d \phi &= 
\left({\delta_{a}}^c + \frac{\nabla_a \phi \nabla^c \phi}{2X} \right) 
\left( \nabla _c \nabla _b \phi + \frac{\nabla_b \phi \nabla^d \phi 
\nabla _c \nabla _d \phi}{2X}\right)\\
&= \nabla _a \nabla _b \phi +  \frac{\nabla_b \phi \nabla^d \phi \nabla_a  
\nabla _d \phi}{2X}
+\frac{\nabla_a \phi \nabla^c \phi \nabla _c \nabla _b \phi}{2X}
+ \frac{\nabla_a \phi \nabla^c \phi \nabla_b \phi \nabla^d \phi \nabla _c 
\nabla _d \phi}{4X^2}\\
&=
\nabla _a \nabla _b \phi +  \frac{\nabla_b \phi \nabla^d \phi \nabla _a 
\nabla _d \phi}{2X}
+\frac{\nabla_a \phi \nabla^c \phi \nabla _b \nabla _c \phi}{2X}
+ \frac{\nabla_a \phi \nabla_b \phi \nabla^c \phi \nabla^d \phi \nabla _c 
\nabla _d \phi}{4X^2}\\
&= \nabla _a \nabla _b \phi 
- \frac{\nabla_{(a} X \nabla_{b)} \phi}{X}
- \frac{\nabla X \cdot \nabla \phi}{4X^2} \, \nabla_a \phi \nabla_b \phi 
\,.
\end{split}
\ee

\section{Computation of ${h_a}^c {h_b}^d \Box\phi \nabla_c\nabla_d\phi$ 
and $ {h_a}^c {h_b}^d \nabla^e\nabla_c \phi \nabla_e\nabla_d\phi$}
\label{sec:AppendixB}
\renewcommand{\theequation}{B.\arabic{equation}}
\setcounter{equation}{0}

Here we compute two terms needed for the evaluation of the effective 
stress tensor~(\ref{chimelafattofare}). We have
\begin{eqnarray}
{h_a}^c {h_b}^d \Box\phi \nabla_c\nabla_d\phi &=& \Box\phi 
\left( \delta_a^c +\frac{\nabla_a\phi \nabla^c\phi}{2X} \right) 
\left( \delta_b^d +\frac{\nabla_b\phi \nabla^d\phi}{2X} \right) 
\nabla_c\nabla_d\phi \nonumber\\
&&\nonumber\\
&=& \Box\phi \left[ \nabla_a \nabla_b \phi -\frac{1}{2X} \left( 
\nabla_a\phi\nabla_b X + \nabla_b\phi\nabla_a X \right)
-\frac{1}{4X^2} \nabla_a\phi\nabla_b\phi \nabla^c\phi\nabla_c\phi \right] 
\,. \label{cazzuto1}
\end{eqnarray}

We then need
\begin{eqnarray}
{h_a}^c {h_b}^d \nabla^e\nabla_c \phi \nabla_e\nabla_d\phi &=& 
\left( \delta_a^c +\frac{\nabla_a\phi\nabla^c\phi}{2X} \right)
\left( \delta_b^d +\frac{\nabla_b\phi\nabla^d\phi}{2X} \right) 
\nabla_e\nabla_c \phi \nabla^e\nabla_d\phi \nonumber\\
 &&\nonumber\\
&=& \nabla_e\nabla_a\phi \nabla^e\nabla_b \phi +\frac{1}{2X} \left[
\nabla_e\nabla_a\phi \nabla^e \nabla_d \phi \nabla_b\phi \nabla^d\phi 
+\left( \nabla_e\nabla_c \phi +\left( \nabla_e\nabla_c \phi \nabla^e 
\nabla_b\phi \right) \nabla_a\phi \nabla^c \phi \right] \right.\nonumber\\
&&\nonumber\\
&\, & \left. +\frac{1}{4X^2} \, \nabla_a\phi \nabla_b\phi \left( \nabla^c 
\phi 
\nabla^d \phi \nabla_e\nabla_c \phi \nabla^e \nabla_b \phi \right) 
\nabla_a \phi \nabla^c \phi \right] \nonumber\\
&&\nonumber\\
&=& \nabla_e\nabla_a\phi \nabla^e\nabla_b \phi -\frac{1}{2X} \left[
\nabla_e\nabla_a\phi \nabla^eX  \nabla_b\phi
+ \left(  \nabla_e X \nabla^e \nabla_b \phi \right) \nabla_a \phi \right]
+ \frac{ \nabla_e X \nabla^e X \nabla_a \phi \nabla_b \phi}{4X^2} 
\,.\label{cazzuto2}
\end{eqnarray}

\medskip

\section*{Acknowledgments}

A.G.~is supported by the European Union's 
Horizon 2020 research and innovation programme under the Marie Sk\l{}odowska-Curie Actions (grant 
agreement No.~895648--CosmoDEC). The work of A.G~has also been carried out in the framework of 
activities of the Italian National Group of Mathematical Physics 
[Gruppo Nazionale per la Fisica Matematica (GNFM), Istituto Nazionale di Alta Matematica (INdAM)].
L.H.~is supported by funding from the European Research Council (ERC) under the European
Union's Horizon 2020 research and innovation programme grant agreement No.~801781 
and by the Swiss National Science Foundation grant No.~179740.
V.F. is supported by the Natural Sciences \& Engineering Research Council 
of Canada (Grant 2016-03803).

\end{document}